%% file: axion.tex
\documentclass[aps,prx,twocolumn,longbibliography,floatfix,10pt,superscriptaddress]{revtex4-1}
\usepackage[utf8]{inputenc}

\usepackage{natbib}
\usepackage{graphicx}
\usepackage{latexsym}
\usepackage{amssymb}
\usepackage{amsmath}
\usepackage{amsfonts}
\usepackage{gensymb}
\usepackage{bm}
\usepackage{hyperref}
\usepackage{braket}
\usepackage{physics}
\usepackage{bbold}
\usepackage[caption=false]{subfig}
\usepackage[dvipsnames]{xcolor}
\usepackage[normalem]{ulem}
\usepackage{siunitx}
\usepackage{mhchem}
\usepackage{multirow}

\definecolor{myforestgreen}{RGB}{34,139,34}

\newcommand{\supplementarysection}{%
  \setcounter{figure}{0}
  \let\oldthefigure\thefigure
  \renewcommand{\thefigure}{S\oldthefigure}
  \setcounter{section}{0}
  \let\oldthesection\thesection
  \renewcommand{\thesection}{S\oldthesection}
  \setcounter{equation}{0}
  \let\oldtheequation\theequation
  \renewcommand{\theequation}{S\oldtheequation}
  \setcounter{table}{0}
  \let\oldthetable\thetable
  \renewcommand{\thetable}{S\oldthetable}
}

\newcommand{\bi}{\begin{itemize}}
\newcommand{\ei}{\end{itemize}}
\newcommand{\be}{\begin{enumerate}}
\newcommand{\ee}{\end{enumerate}}
 
\newcommand{\Eq}[1]{Eq.(\ref{#1})}
\newenvironment{dfn}{{\vspace*{1ex} \noindent \bf Definition }}{\vspace*{1ex}}

\newcommand{\nn}{\nonumber}  %

	\newcommand{\beq}{\begin{eqnarray}}
	\newcommand{\eeq}{\end{eqnarray}}

\begin{document}

\title{Detecting axion dynamics on the surface of magnetic topological insulators}

\author{Zhi-Qiang Gao}
\affiliation{Department of Physics, University of California, Berkeley, CA 94720, USA \looseness=-2}
\affiliation{Material Science Division, Lawrence Berkeley National Laboratory, Berkeley, CA 94720, USA \looseness=-2}
\author{Taige Wang}
\affiliation{Department of Physics, University of California, Berkeley, CA 94720, USA \looseness=-2}
\affiliation{Material Science Division, Lawrence Berkeley National Laboratory, Berkeley, CA 94720, USA \looseness=-2}
\affiliation{Kavli Institute for Theoretical Physics, University of California, Santa Barbara, CA 93106, USA \looseness=-2}
\author{Michael P. Zaletel}
\affiliation{Department of Physics, University of California, Berkeley, CA 94720, USA \looseness=-2}
\affiliation{Material Science Division, Lawrence Berkeley National Laboratory, Berkeley, CA 94720, USA \looseness=-2}
\author{Dung-Hai Lee}
\affiliation{Department of Physics, University of California, Berkeley, CA 94720, USA \looseness=-2}
\affiliation{Material Science Division, Lawrence Berkeley National Laboratory, Berkeley, CA 94720, USA \looseness=-2}


\begin{abstract}

Axions, initially proposed to solve the strong CP problem, have recently gained attention in condensed matter physics, particularly in topological insulators. However, detecting axion dynamics has proven challenging, with no experimental confirmations to date. In this study, we identify the surface of magnetic topological insulators as an ideal platform for observing axion dynamics. The vanishing bulk gap at the surface allows for order $O(1)$ variations in the axion field, making the detection of axion-like phenomena more feasible. In contrast, these phenomena are strongly suppressed in the bulk due to the small magnetic exchange gap. We investigate two-photon decay as a signature of axion dynamics and calculate the branching ratio using a perturbative approach. Our findings reveal that the photon flux emitted from the surface is in-plane and orders of magnitude larger than that from the bulk, making it detectable with modern microwave technology. We also discuss potential material platforms for detecting axion two-photon decay and strategies to enhance the signal-to-noise ratio.

\end{abstract}

\maketitle

\textbf{Introduction.} Axions were first introduced in the 1980s to address the strong CP problem and later became a candidate for dark matter \cite{Peccei1977,Wilczek1978,Weinberg1978}. Despite extensive efforts to detect axions in the universe \cite{Asztalos2010,admx}, they have yet to be observed to date.

In the past decade, the concept of the axion has found a new application in condensed matter physics, particularly in the study of topological insulators. Many materials exhibit magnetoelectric responses, as exemplified by phenomena like Faraday and Kerr rotation \cite{Sekine2021}. These responses can generally be described by the bulk action:
\begin{equation} \label{eq:theta}
\mathcal{L}_{\theta}^B=\frac{\alpha}{\pi}\theta\mathbf{E}\cdot\mathbf{B},
\end{equation}
where $\theta$ is defined modulo $2\pi$, $\mathbf{E}$ and $\mathbf{B}$ are the electric and magnetic fields, respectively, and $\alpha$ is the fine-structure constant. While $\theta$ is typically small for most materials, in topological insulators, it is large and quantized at $\theta = \pi$ \cite{Qi2008}. This action arises from the Berry phase associated with the chiral transformation parameterized by $\theta$ \cite{DHL2022}. Moreover, unless protected by symmetry, the coefficient $\theta$ can take any value and fluctuate in both space and time, $\theta(\mathbf{r},t)$ \cite{Li2010}. 
These fluctuations can be interpreted as an axion field because of the same coupling to the electromagnetic field.

Detecting axion dynamics in experiments has been challenging. Various experimental signatures have been proposed, such as the emergence of axionic polaritons and the dynamical chiral magnetic effect \cite{Li2010}, but none have yet been observed. In this letter, we argue that the surface of magnetic topological insulators presents an ideal platform for detecting axion dynamics. In these materials, the value of $\theta$ is determined by two energy scales: the bulk gap $m_0$ and the surface exchange gap $m_a$ induced by magnetic order. Within a four-band Dirac model, it can be explicitly shown \cite{Sekine2021},
\begin{equation} \label{eq:Fujikawa}
\theta=\frac{\pi}{2}\left[1-\operatorname{sgn}\left(m_0\right)\right]-\tan^{-1}\left(\frac{m_a}{m_0}\right).
\end{equation}
At temperatures below $m_0$ but above $m_a$, the charge degrees of freedom are frozen in the bulk, and only the spin degrees of freedom are active. Consequently, the dynamics of $m_a$ can be viewed as the axion dynamics of the $\theta$ field via Eq.~\ref{eq:Fujikawa}.

In most magnetic topological insulators, including the $\operatorname{MnTe}\left(\mathrm{Bi}_2 \mathrm{Te}_3\right)_n$ \cite{Otrokov2019} and $\mathrm{EuX}_2\mathrm{As}_2$ families \cite{EuInAs,EuSnAs}, the surface exchange gap $m_a \sim \SI{1}{meV}$ \cite{MBTARPES,MBTARPES2} is much smaller than the bulk gap $m_0 \sim \SI{100}{meV}$ \cite{Wang2020aa}. This ratio significantly constrains the fluctuation of the $\theta$ field since $\Delta\theta \sim m_a/m_0 \ll 1$. Even if $m_a$ fluctuates by an order of magnitude comparable to its average value, the resulting fluctuation in $\theta$ is only $\delta\theta \sim 10^{-2}$, making it nearly undetectable. Although recent first-principles searches have extensively explored materials with larger $m_a$ \cite{Wang2020aa}, $m_a$ in the bulk cannot exceed $m_0$; otherwise, charge and spin degrees of freedom would mix, significantly altering the low-energy behavior.

However, at the surface of magnetic topological insulators, the bulk gap $m_0$ must close because the vacuum is topologically distinct from the bulk. Under such conditions, \Eq{eq:theta} is replaced by the surface action
\beq \label{CS}
\mathcal{L}^S_{\theta}=\frac{\alpha}{\pi}\theta\epsilon^{\mu\nu\lambda}A_\mu\partial_\nu A_\lambda.
\eeq
Here $\theta=\pi~{\rm sgn}(m_a)$. As a result, even small fluctuations in $m_a$ at the surface can lead to $O(1)$ fluctuations in the $\theta$ field, making axion dynamics much easier to detect. To illustrate that the surface of magnetic topological insulators is indeed an ideal platform for detecting axion dynamics, we examine a specific experimental consequence—two-photon decay. According to Eq.~\ref{eq:theta}, an axion can decay into two photons, which is one of the primary methods proposed to detect axion dark matter in the universe. In contrast, in materials, axions can also decay into various low-energy excitations, such as excitons and phonons, due to the underlying many-body interaction. These decay channels often dominate over the photon channel. The main result of this letter is an estimated branching ratio—the fractional probability—that axions decay into two photons compared to other excitations.

We consider a concrete setup where two-photon decays are induced by the flip of magnetic domains and calculate the contributions from both the bulk and the surface. It turns out that the branching ratio scales as $\delta \theta^2$, and therefore the contribution from the surface is $\sim 10^4$ times larger than in the bulk. For a macroscopic sample, this results in a photon flux detectable with modern microwave technologies. Finally, we discuss strategies to enhance signal-to-noise ratio, such as stimulated emission by a photon background in the environment.

\begin{figure}[htbp]
     \centering
     \includegraphics[width=0.95\linewidth]{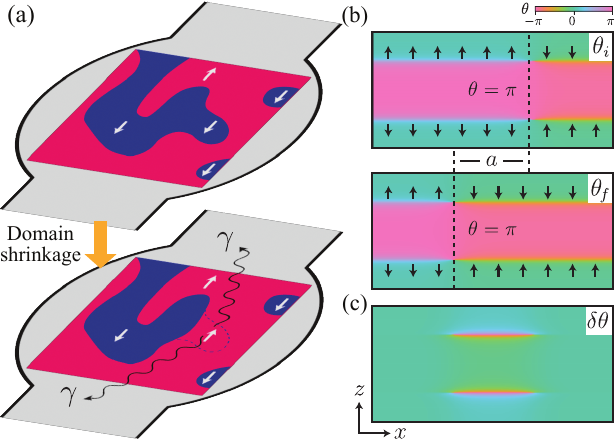}
     \caption{(a) Schematics of the experimental setup. The magnetic topological insulator is placed in a microwave cavity. Spins on the \textit{top} surface point up in the red region and down in the blue region. When spin flip within the dashed line enclosed region, there is a finite probability that two in-plane photons will be emitted through the two-photon decay channel of axions. (b-c) A side view of the spin configuration before and after the domain wall movement, and the corresponding $\theta$ field. $\theta$ only changes a significant amount at the surface.}
     \label{fig:dw}    
 \end{figure}

\textbf{Axion decay from domain shrinkage.} Domain walls naturally form on the surface of magnetic topological insulators below the Néel temperature, as evidenced by magnetic imaging \cite{nanolett}. These domains exhibit significant mobility, disappearing rapidly as the temperature approaches the Néel temperature. An applied magnetic field can cause domains with opposite magnetization to shrink. Additionally, magnetic domains can be manually created by shining circular polarized light on the surface \cite{Qiu2023}. Deep inside the bulk, $\theta$ is nearly quantized to $\pi$. However, on the surface, $\theta=\pi~\mathrm{sgn}(m_a)$. Hence upon spin reversal $\theta$ can wind from $\pm \pi$ (which corresponds to a surface Hall conductance $\pm e^2/2h$) to $\mp\pi$ through $0$ depending on the magnetization of the domain . 

Fig.~\ref{fig:dw} (a) (with the side view in (b)) depicts a snapshot of the spin configurations on the surface of a magnetic topological insulator. When spins flip within the region enclosed by the dashed lines, the $\theta$ field correspondingly changes from $\theta_i=-\pi$ to $\theta_f=\pi$. As shown in Fig.~\ref{fig:dw} (b-c), this change in the $\theta$ field is significant only at the surface, while within the bulk, the change is at the order of $m_a/m_0$. Here, the dynamics of the $\theta$ field is tied to the dynamics of the magnetic moments. From a spin perspective, axion decay can also be viewed as an additional, radiative, dissipation channel for spin dynamics, driven by the coupling of the $\theta$-term to the electromagnetic fields. In particular, the change in the magnetic energy associated with the magnetic domain reversal can be carried away by the emitted photons.

The absolute decay rate depends on the microscopic details of the dominant decay channels other than the two-photon channel. In the following, we show that without knowing the specific details of such decay channels, it is still possible to calculate the branching ratio $\eta$, namely, the fractional decay rate, of the two-photon decay channel. 

We begin by introducing the Hamiltonian $ H_1 $, which governs the dynamics of the axion field $ \delta\theta $, and treat the interaction between the axion field and the electromagnetic field, represented by $ H_2 $, as a perturbation. Working in the interaction picture with respect to the Hamiltonian for the free electromagnetic field $ H_0^\mathrm{EM} $, the full system can be described by:
\begin{equation} \label{eq
}
H=H_0^\mathrm{EM}+H_1[\delta\theta(\mathbf{r})]+H_2(t).
\end{equation}
Since $ H_1 $ does not couple to electromagnetic fields, it remains unaffected in the interaction picture defined by $ H_0^\mathrm{EM} $. It is notably that $ H_1 $ determines the dominant decay channels since it includes the coupling between the axion field and excitations such as phonons and excitons. The precise form of the coupling term $ H_2 $ depends on whether the interaction occurs in the bulk \Eq{eq:theta} or at the surface \Eq{CS}.

The transition amplitude of the two-photon decay process over a time interval $T$ is given by:
\beq
G(T;\mathbf{k},\mathbf{k}^\prime)=\bra{\delta\theta_f(\mathbf{r}),\mathbf{k},\mathbf{k}^\prime}\mathcal{T}e^{-i\int_{0}^{T} \mathrm{d}t (H_1+H_2(t))}\ket{\delta\theta_i(\mathbf{r})},
\label{eq:G}
\eeq
where $\mathbf{k}$ and $\mathbf{k}^\prime$ are the momenta of the emitted photons, and $\delta\theta_i(\mathbf{r})$ and $\delta\theta_f(\mathbf{r})$ are the initial and the final axion field configurations. Now, considering  spin flips in a region of linear size $a$ (see Fig.~\ref{fig:dw} (b) for a side view), on length scale much greater than $a$ the $\delta\theta$ field can be approximated as $\delta\theta_i(\mathbf{r}) \sim \theta_0 \delta(\mathbf{r}-\mathbf{r}_0)$ and $\delta\theta_f(\mathbf{r})=0$, where $\mathbf{r}_0$ is the center of the spin flip region. We can evaluate $G(T;\mathbf{k},\mathbf{k}^\prime)$ using standard time-dependent perturbation theory to first order in $H_2$ (see S.M. Section I for details \cite{supp}):
\beq
G(T;\mathbf{k},\mathbf{k}^\prime)&=&\frac{-i\alpha}{2\pi}G_0(T)f(\mathbf{k},\mathbf{k}^\prime)\nonumber \\
&\times & \int_{\mathbf{r},t}\left<\delta\theta(\mathbf{r},t)\right>_0e^{i(\mathbf{k}+\mathbf{k}^\prime)\cdot\mathbf{r}-i(\omega_\mathbf{k}+\omega_{\mathbf{k}^\prime})t}.
\eeq
Here, $f(\mathbf{k},\mathbf{k}^\prime)$ is the form factor that depends on $H_2$, and $\omega_\mathbf{k}=c|\mathbf{k}|$ is the photon dispersion. $G_0(T)$ and $\left<\delta\theta(\mathbf{r},t)\right>_0$ represent the transition amplitude and the $\theta$ field evolution in the absence of $H_2$. Since $H_2(t)$ is diagonal in $\theta$, the decay of $\delta\theta$ is governed entirely by $H_1$.


Although the exact form of $\left<\delta\theta(\mathbf{r},t)\right>_0$ is unknown due to the dependence of $H_1$ on microscopic details, its qualitative behavior can be described using a dissipative two-level system \cite{Leggett1987, Wu2007}. The Ising-like ferromagnetic ordering at the surface creates two minima in the $\theta$ field, corresponding to $\theta = \pm \pi$, which we treat as discrete levels \cite{Wu2007}. During the magnetic transition, the $\theta$ field moves from the upper level, $\theta_i = -\pi$, to the lower level, $\theta_f = \pi$. The dynamics of two-level systems coupled to a thermal bath have been extensively studied \cite{Leggett1987, Wu2007}, and the form of $\left<\delta\theta(\mathbf{r},t)\right>_0$ is determined by the spectral function of the bath. However, in the transition regime where flips between levels occur, $\left<\delta\theta(\mathbf{r},t)\right>_0$ has two important features: it \textit{decays exponentially} with some rate $\Gamma_0$; and it, optionally, \textit{oscillates} at frequency $\Delta$ equal to the energy difference between the two levels. Here $\Gamma_0=\lim_{T\to +\infty}|G_0(T)|^2/T$ is the bare decay rate in the absence of $H_2$, and $\Delta$ measures the energy difference gained by spin flips within the region of linear size $a$. Since the two-photon channel decay rate $\Gamma$ depends only on the exponential decay, we approximate $\left<\delta\theta(\mathbf{r},t)\right>_0$ as $\delta\theta_i(\mathbf{r}) e^{-\Gamma_0 t}$. Then we find
\begin{equation}
    \Gamma \equiv \lim_{T\to\infty}\int_{\mathbf{k},\mathbf{k}^\prime}\frac{|G(T;\mathbf{k},\mathbf{k}^\prime)|^2}{T} = C\theta_0^2\alpha^2\Gamma_0
\end{equation}
up to the leading order of $\Gamma_0 a/\hbar c$, where the dimensionless number $C$ depends on whether the contribution is from the bulk or the surface. In the following, we plug the axion relaxation into \Eq{eq:theta} or \Eq{CS} to determine the exact branching ratio.

Nevertheless, it is already clear that the surface contribution is significantly larger, as $\theta_0$ is on the order of one at the surface but on the order of $m_a/m_0 \sim 10^{-2}$ deep inside the bulk.

\textbf{Surface and bulk contribution.} In the radiation gaue ($A_0=0$), at the surface of magnetic topological insulators, the coupling between the $\theta$ field and the electromagnetic field takes the form of a Chern-Simons (CS) term:
\begin{equation}
H_2^S(t)=\frac{\alpha}{\pi}\int\mathrm{d}^2\mathbf{r}~\delta\theta(\mathbf{r})\left(\mathbf{A}(\mathbf{r},t)\times\partial_t\mathbf{A}(\mathbf{r},t)\right)\cdot \hat{\mathbf{z}}.\label{eq:Hsurface}
\end{equation}
Here, we consider the surface at $z = 0$, so the CS term involves only the in-plane components $A_x$ and $A_y$ of the vector potential. Consequently, the momentum of the photons emitted during domain wall shrinkage must also be in-plane, as the radiated electromagnetic field is transverse. Intuitively, the electric field $\mathbf{E}$ lies in-plane, and since the current is also in-plane, the magnetic field $\mathbf{B}$ must be out-of-plane, resulting in an electromagnetic wave with in-plane momentum in the direction of  $\mathbf{E} \times \mathbf{B}$. Therefore, the combination of a 2D cavity and waveguide, as shown in Fig.~\ref{fig:dw}(a), is an efficient setup for detecting these in-plane photons.

In S.M. Section I, we compute the corresponding form factor $f^S(\mathbf{k},\mathbf{k}^\prime)$ for the CS term, and the resulting decay rate is:
\beq
\Gamma^S = 0.006\theta_0^2\alpha^2\Gamma_0,
\eeq
yielding a branching ratio $\eta^S = 0.006\theta_0^2\alpha^2 \sim 10^{-5}$ for $\theta_0\approx 2\pi$.

One might be concerned whether the gapless modes at the magnetic domain wall contribute to photon emission as well. In S.M. Section II \cite{supp}, we derive the effective action for the electromagnetic field after integrating out the domain wall modes \cite{Wen1992}:
\beq
S_{\mathrm{DW}}=-\frac{1}{4 \pi} \sum_{\omega, k} A(-\omega,-k) \frac{\omega^2}{k(i \omega-v k)} A(\omega, k),
\eeq
where $k$ is the 1D momentum along the domain wall, $A$ is the component of the vector potential along the domain wall, and $v$ is the edge velocity of the gapless modes. Since the gauge field is longitudinal and its propagator strongly peaked at $k=0$, we conclude that the gapless modes living on the domain wall do not emit any finite-frequency photons.

Now, we switch to the bulk contribution, where:
\begin{equation}
    H_2^B(t)=\frac{\alpha}{\pi}\int\mathrm{d}^3\mathbf{r}~\theta(\mathbf{r})\mathbf{E}(\mathbf{r},t)\cdot\mathbf{B}(\mathbf{r},t).\label{eq:Hsf}
\end{equation}
The branching ratio is given by (see S.M. Section III for details):
\begin{equation}
    \eta^B=0.02\theta_0^2\alpha^2.
\end{equation}
Although the prefactor is larger than that for the surface, we note that $\theta_0 \sim m_a/m_0 \sim 10^{-2}$ in the bulk, leading to a branching ratio of only $\eta \sim 10^{-9}$. The branching ratio we obtain here is order of magnitude similar to that of long-wavelength axion (magnon) decay computed using standard field theory techniques (see S.M. Section IV \cite{supp} for details).

The decay rate is independent of the size $a$ of the domain being flipped. However, the number of domain shrinkage events $N$ (which we referred to as the axion number below) determines the total photon flux from the branching. We estimate the upper bound of $N$ by considering the minimum size of each domain, given by the correlation length of the underlying magnetic topological insulator, with $a \sim \xi \sim \SI{1}{nm}$ for most materials \cite{Sekine2021}. On the other hand, the sample size is typically constrained to $L \sim \SI{1}{\mu m}$ in most experiments. Combining these limits, the total number of axions on the surface is $N^S \sim 10^6$, and in the bulk with thickness of the sample $d \sim \SI{10}{nm}$ gives $N^B \sim 10^7$.  To obtain one photon on average during the magnetic transition of the entire sample, the branching ratio must be $\eta^B > 10^{-7}$ and $\eta^S > 10^{-6}$. From previous calculations, we conclude that the total photon flux from the surface is at least three orders of magnitude higher than that from the bulk, indicating that the surface is a far better location to observe axion dynamics. At the surface, the total number of emitted photons after all the spin are reversed is around 10 for a typical sample, which should be detectable by single-photon detectors. For the bulk signal to be detectable, $\theta_0$ should be at least $O(10^{-1})$ for a typical sample, which might be achievable by driving the material close to the topological transition point with a smaller bulk gap $m_0$.

The main experimental challenge is to distinguish photons emitted from axion decay from those existed in the environment. It is advantageous to have spin flipping occur within a relatively short time interval, which can be achieved in materials with strong spin-exchange coupling. For instance, with a spin-flipping time on the order of one second, the luminosity of the emitted photons would be around 10 photons per second.

\begin{figure}[htbp]
     \centering
     \includegraphics[width=0.9\linewidth]{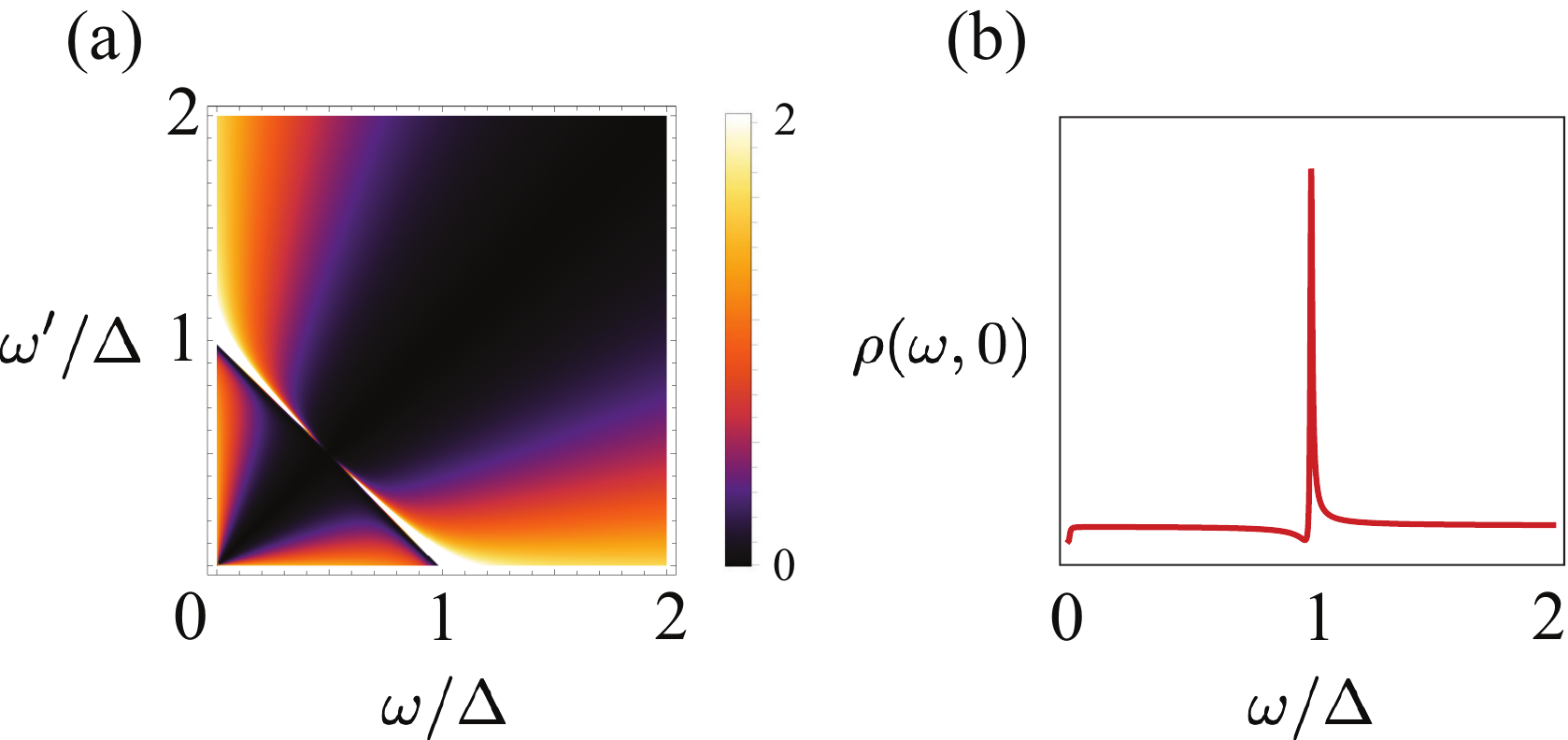}
     \caption{The energy distribution of the emitted photons in the domain wall decay setup. $\omega$ and $\omega^\prime$ represent the energies of the two emitted photons, and $\Delta$ is the reduction in exchange energy during axion decay. The peak at $\omega+\omega^\prime=\Delta$ is very sharp. (b) is plotted along the $\omega^\prime=0$ line, i.e. the $x$-axis in (a).}
     \label{fig:dos}    
 \end{figure}

As discussed before, the axion relaxation contains an oscillatory piece, whose frequency is equal to $\Delta$ and amplitude much smaller than 1. In the presence of $\Delta$, the energy distribution (density of states) of the emitted photons is shown in Fig.~\ref{fig:dos}, which peaks at $\omega+\omega^\prime=\Delta$ with a broadening $\Gamma_0 \ll \Delta$, reflecting the energy conservation. However, the energy conservation is not strict since the peak is not a $\delta$-function. This is attributed to the fact that axion can exchanges energy with excitations in other decay channels (see S.M. Section I \cite{supp} for a detailed calculation).

\textbf{Stimulated emission.} One way to distinguish photons from axion decay and the environment is through a locked-in experiment using stimulated emission. If we shine microwave light at a particular frequency $\omega_0$ at the sample, spins can absorb the light and re-emit at frequency $\Omega = \Delta + \omega_0$. Though $\Delta$ might not be sharply defined, it is possible to track the peak frequency of the emitted photons as a function of $\omega_0$, providing a more convincing evidence of axion decay.

\begin{figure}[htbp]
     \centering
     \includegraphics[width=0.7\linewidth]{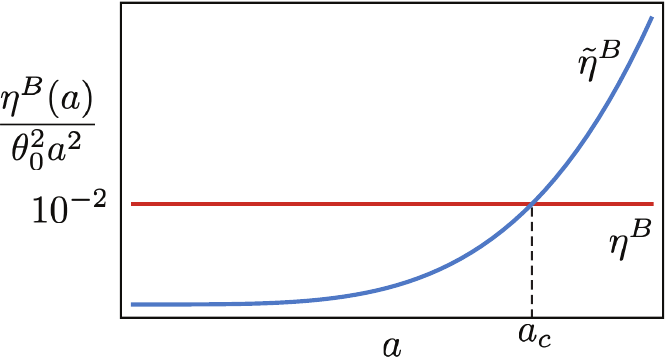}
     \caption{The bulk branching ratios of the photon channel in the domain wall shrinkage setup as functions of the axion size $a$. The spontaneous two-photon emission branching $\eta^B$ is independent of $a$. The stimulated emission branching $\tilde{\eta}^B$ is proportional to $a^4$. The branching of the stimulated emission is larger $\tilde{\eta}^B > \eta^B$ only when $a>a_c$.}
     \label{fig:stimulate}    
 \end{figure}

For the bulk, the branching in the presence of the stimulated emission is (see S. M. Section V A for details \cite{supp})
\beq
\tilde{\eta}^B\approx 10^{-5}\times\frac{Ia^4\theta_0^2}{\hbar c^2},
\eeq
up to leading order in $\omega_0 a/c$, where $I=\hbar cn\omega_0$ is the power density of the stimulating photons with photon density $n$ and frequency $\omega_0$. 
In Fig.~\ref{fig:stimulate} we show the branching ratios of the photon channel with and without stimulation as functions of the axion size $a$. When the axion size exceeds the critical value $a_c\approx\sqrt[4]{\hbar c^2/I}\approx \SI{1}{\mu m}$ for $I\approx \SI{1}{kW/cm^2}$, the branching will be comparable to the spontaneous decay setup $\tilde{\eta}^B>\eta^B$. However, in practice, the axion size is much smaller and therefore we expect spontaneous decay to be dominating over stimulated emission for the bulk axion decay.

For the surface, the branching in the presence of stimulating photon reads (see S. M. Section V B for details \cite{supp})
\beq
\tilde{\eta}^S\approx 10^{-5}\times\nu a^2 \theta_0^2,
\eeq
where $\nu$ is the 2D photon density. In contrast to the bulk contribution, the branching at the surface can, in principle, be greatly enhanced compared with the spontaneous decay case, if the stimulating photon number per flipped domain $\nu a^2$ can exceed $O(1)$.

\textbf{Experimental considerations.} We first discuss possible material platforms to detect axion decay. As we mentioned previously, magnetic topological insulators are a promising candidate, including the $\operatorname{MnTe}\left(\mathrm{Bi}_2 \mathrm{Te}_3\right)_n$ family, and the $\mathrm{EuX}_2\mathrm{As}_2$ family. Both of them are layer materials with intra-layer ferromagnetic order and inter-layer antiferromagnetic order, and $\theta$ is expected to be close to $\pi$ in the bulk. Since the surface is most important for the two-photon decay, we only need to focus on the domain walls on the top layer. In other words, spins on the top and bottom layer do not have to anti-align as suggested by Fig.~\ref{fig:dw} (b). The magnetic moments also do not have to be intrinsic to the material. An alternative material candidate is (Fe,Cr)-doped (Bi,Sb)$_2$(Se,Te)$_3$ \cite{Chang2013,Wu2016,Okada2016}, in which the magnetic dopants play the role of magnetic moments on the top and bottom surfaces. 

In experiments, domain walls can be generated through optical induction by pump light \cite{Qiu2023}. The decay of these light-induced domain walls will be resulted in the emission of photons with energy comparable with the domain wall tension times the axion size, i.e. the extra energy caused by a unfavorable domain of size $a$. Typically this AFM spin exchange energy in MnBi$_2$Te$_4$ family is at the order of 1 meV. Therefore the corresponding wavelength of the emitted photon is around 1 mm, which falls into the microwave region. For the Fe or Cr doped Bi$_2$(Se,Te)$_3$, the FM spin exchange energy can be $\sim$ 10 meV, and the wavelength of the emitted photons may reach the infrared region.

In conclusion, through computing the branching ratio of two-photon emission versus other spin relaxation channels, we find the reversal of magnetic domains on the surface magnetic topological insulator is a far better candidates for detecting axion dynamics. This invites the revisit of various proposed axion dynamics experiments to include the surface contributions. In addition, our estimates suggest the emitted photon is within the reach of modern single-photon microwave detector.

\textbf{Acknowledgments.} We thank Zhi-Xun Shen, Zhurun Ji, Yayu Wang, Yuanbo Zhang, Suyang Xu, and Salvatore Pace for helpful discussions. This work was funded by the U.S. Department of Energy, Office of Science, Office of Basic Energy Sciences, Materials Sciences and Engineering Division under Contract No. DE-AC02-05-CH11231 (Theory of Materials program KC2301). T.W. is also supported by the Heising-Simons Foundation, the Simons Foundation, and NSF grant No. PHY-2309135 to the Kavli Institute for Theoretical Physics (KITP).

\bibliography{axion}

\newpage

\onecolumngrid

\vspace{0.3cm}

\supplementarysection

\input{supp.tex}

\vfill 

\end{document}

%% file: supp.tex
\begin{center}
\Large{\bf Supplemental Material for ``Detecting axion dynamics on the surface of magnetic topological insulators"}
\end{center}

For clarity in the Supplemental Material we use $\theta(\mathbf{r},t)$ to represent the deviation of the axion field from $\pi$ or $-\pi$, which is denoted as $\delta\theta(\mathbf{r},t)$ in the main text. The unit $\hbar =c=k_B=\epsilon_0 =1$ is employed in the derivation without further notices.

\section{Axion decay from surface domain wall shrinkage}

On the surface, the axion field and the electromagnetic field coupled through a Chern-Simons (CS) term. Without losing of generality, we assume the surface plane is perpendicular to the $z$-axis. Correspondingly the Hamiltonian reads
\beq
&&H=H_0^\mathrm{EM}+H_1[\theta]+H_2^S(t),\nn\\
&&H_2^S(t)=\frac{\alpha}{\pi}\int\mathrm{d}^2\mathbf{r}~\theta(\mathbf{r})\left(A_y(\mathbf{r},t)\partial_t A_x(\mathbf{r},t)-A_x(\mathbf{r},t)\partial_t A_y(\mathbf{r},t)\right).\label{eq:Hsf}
\eeq 
The transition amplitude describing the two-photon emission process in a time interval $T$ is
\beq
G(T;\mathbf{k},\mathbf{k}^\prime)=\bra{\theta_f(\mathbf{r}),\mathbf{k},\mathbf{k}^\prime}\mathcal{T}\exp\left(-i\int_{0}^{T} \mathrm{d}t~H_1[\theta]+H^S_2(t)\right)\ket{\theta_i(\mathbf{r})},
\eeq
where $\theta_i(\mathbf{r})$ and $\theta_f(\mathbf{r})$ are the initial and final field configurations of the axion field with $\theta_f(\mathbf{r})=0$ for all $\mathbf{r}$, as stated in the main text. The states of the emitted photons are labeled by their momentum $\mathbf{k}$ and $\mathbf{k}^\prime$ (and polarization $\lambda$ and $\lambda^\prime$, omitted for clarity), with normalization
\beq
\ket{\mathbf{k}}=\sqrt{(2\pi)^3 2|\mathbf{k}|}a^\dagger(\mathbf{k})\ket{0}.
\eeq
Here $a^\dagger(\mathbf{k})$ creates a photon mode with momentum $\mathbf{k}$ (polarization omitted for clarity), satisfying commutation relation $[a(\mathbf{k}),a^\dagger(\mathbf{k}^\prime)]=\delta^{(3)}(\mathbf{k}-\mathbf{k}^\prime)$. The mode expansion of the electromagnetic field reads
\beq
\mathbf{A}(\mathbf{r},t)=\int\frac{\mathrm{d}^3\mathbf{k}}{\sqrt{(2\pi)^3 2|\mathbf{k}|}}\sum_{\lambda=1,2}\mathbf{e}_\lambda(\hat{\mathbf{k}})a_\lambda(\mathbf{k})e^{i|\mathbf{k}|t-i\mathbf{k}\cdot\mathbf{r}}+\mathbf{e}^*_\lambda(\hat{\mathbf{k}})a_\lambda^\dagger(\mathbf{k})e^{-i|\mathbf{k}|t+i\mathbf{k}\cdot\mathbf{r}},\label{eq:A}
\eeq
where $\mathbf{e}_\lambda(\hat{\mathbf{k}})$, with $\hat{\mathbf{k}}=\frac{\mathbf{k}}{|\mathbf{k}|}$, is the polarization vector of the photon with polarization $\lambda$ and momentum $\mathbf{k}$. The photon dispersion $\omega_\mathbf{k}=|\mathbf{k}|$ is used. Note that while the axion field $\theta(\mathbf{r},t)$ is defined on the two-dimensional surface, the electromagnetic field still lives in the three-dimension space. Thus, the photon momentum $k$ still has three components. However, since the surface axion field only couples to $A_x$ and $A_y$ components of the gauge potential in \Eq{eq:Hsf}, the transverse-ness of the electromagnetic field requires the photons emitted via the surface axion decay must have momenta parallel to the surface plane. This is reflected in the measure of the phase space, as we will discuss later.

The transition amplitude $G(T;\mathbf{k},\mathbf{k}^\prime)$ can be evaluated by insertion of identity operators
\beq
\mathbb{I}_a=\int D[\theta]\ket{\theta(\mathbf{r})}\bra{\theta(\mathbf{r})},~~\mathbb{I}_{2p}=\frac{1}{2}\int\frac{\mathrm{d}^3\mathbf{k}}{(2\pi)^3 2|\mathbf{k}|}\frac{\mathrm{d}^3\mathbf{k}^\prime}{(2\pi)^3 2|\mathbf{k}^\prime|}\ket{\mathbf{k},\mathbf{k}^\prime}\bra{\mathbf{k,\mathbf{k}^\prime}}.
\eeq
Up to the tree level, the transition amplitude reads
\beq
&&G(T;\mathbf{k},\mathbf{k}^\prime)=\frac{1}{2}\int D[\theta]\int_{t=0}^{T}\int\frac{\mathrm{d}^3\mathbf{q}}{(2\pi)^3 2|\mathbf{q}|}\frac{\mathrm{d}^3\mathbf{q}^\prime}{(2\pi)^3 2|\mathbf{q}^\prime|}\bra{\theta_f(\mathbf{r}),\mathbf{k},\mathbf{k}^\prime}e^{-iH_1[\theta]t}\ket{\theta_{t}(\mathbf{r}),\mathbf{q},\mathbf{q}^\prime}\nn\\
&&\times \bra{\theta_{t}(\mathbf{r}),\mathbf{q},\mathbf{q}^\prime}\left(1-iH^S_2(t)\mathrm{d}t\right)e^{-iH_1[\theta]\mathrm{d}t}\ket{\theta_{t+\mathrm{d}t}(\mathbf{r})}\bra{\theta_{t+\mathrm{d}t}(\mathbf{r})}e^{-iH_1[\theta](T-t-\mathrm{d}t)}\ket{\theta_i(\mathbf{r})}\\
&&=\frac{-i}{2}\int D[\theta]\int\mathrm{d}t\bra{\theta_f(\mathbf{r})}e^{-iH_1[\theta]t}\ket{\theta_{t}(\mathbf{r})}\bra{\theta_{t}(\mathbf{r}),\mathbf{k},\mathbf{k}^\prime}H^\prime(t)e^{-iH_1[\theta]\mathrm{d}t}\ket{\theta_{t+\mathrm{d}t}(\mathbf{r})}\bra{\theta_{t+\mathrm{d}t}(\mathbf{r})}e^{-iH_1[\theta](T-t-\mathrm{d}t)}\ket{\theta_i(\mathbf{r})}\nn,
\eeq
with
\beq
\bra{\theta_{t}(\mathbf{r}),\mathbf{k},\mathbf{k}^\prime}H^S_2(t)e^{-iH_1[\theta]\mathrm{d}t}\ket{\theta_{t+\mathrm{d}t}(\mathbf{r})}=\frac{\alpha}{\pi}\int\mathrm{d}^2\mathbf{r}~\theta(\mathbf{r},t)f^S(\mathbf{k},\mathbf{k}^\prime)e^{i(\mathbf{k}+\mathbf{k}^\prime)\cdot\mathbf{r}-i(|\mathbf{k}|+|\mathbf{k}^\prime|)t}\bra{\theta_t(\mathbf{r})}e^{-iH_1[\theta]\mathrm{d}t}\ket{\theta_{t+\mathrm{d}t}(\mathbf{r})}.
\eeq
Here $f^S(\mathbf{k},\mathbf{k}^\prime)$ is the surface form factor (dependence of the polarizations $\lambda$ and $\lambda^\prime$ is implicit for clarity)
\beq
f^S(\mathbf{k},\mathbf{k}^\prime)=\frac{i}{2}\left(|\mathbf{k}^\prime |-|\mathbf{k}|\right)\left(\mathbf{e}^{*}_\lambda(\hat{\mathbf{k}})\times \mathbf{e}^{*}_{\lambda^\prime}(\hat{\mathbf{k}}^\prime)\right)\cdot\hat{\mathbf{z}}.
\eeq
Note that in the computation of $f^S(\mathbf{k},\mathbf{k}^\prime)$ one needs to symmetrize the CS term coupling in \Eq{eq:Hsf}, since $A_{x,y}$ and $\partial_t A_{y,x}$ does not commute under the canonical quantization \Eq{eq:A}. Thus, the transition amplitude can be further written as
\beq
G(T;\mathbf{k},\mathbf{k}^\prime)&=&\frac{-i\alpha}{2\pi}\int_{\theta(\mathbf{r},0)=\theta_i(\mathbf{r})}^{\theta(\mathbf{r},T)=\theta_f(\mathbf{r})} D[\theta]~e^{iS[\theta]}\int \mathrm{d}^2\mathbf{r}\mathrm{d}t~\theta(\mathbf{r},t)f^S(\mathbf{k},\mathbf{k}^\prime)e^{i(\mathbf{k}+\mathbf{k}^\prime)\cdot\mathbf{r}-i(|\mathbf{k}|+|\mathbf{k}^\prime|)t}\nn\\
&=& \frac{-i\alpha}{2\pi}G_0(T)f^S(\mathbf{k},\mathbf{k}^\prime)\int \mathrm{d}^2\mathbf{r}\mathrm{d}t~\left<\theta(\mathbf{r},t)\right>_0e^{i(\mathbf{k}+\mathbf{k}^\prime)\cdot\mathbf{r}-i(|\mathbf{k}|+|\mathbf{k}^\prime|)t},
\eeq
where
\beq
G_0(T)=\bra{\theta_f(\mathbf{r})}e^{-iH_1[\theta]T}\ket{\theta_i(\mathbf{r})}=\int_{\theta(\mathbf{r},0)=\theta_i(\mathbf{r})}^{\theta(\mathbf{r},T)=\theta_f(\mathbf{r})} D[\theta]~e^{iS[\theta]}
\eeq
is the transition amplitude between field configurations $\theta_i(\mathbf{r})$ and $\theta_f(\mathbf{r})$ controlled by the axion Hamiltonian $H_1[\theta]$. The decay rate of this tunneling is
\beq
\Gamma_0=\lim_{T\rightarrow +\infty}\frac{|G_0(T)|^2}{T},
\eeq
which can be measured in experiments. $\left<\theta(\mathbf{r},t)\right>_0$ is the expectation value of the axion field configuration evaluated under $H_1[\theta]$, which can be approximated by its classical value. Here we simply choose an exponential decay of $\theta(\mathbf{r},t)$ at position $\mathbf{r}=\mathbf{r}_0$ (the small oscillatory piece is temporarily omitted since it does not affect the result at the leading order). Since the photon emission does not induce axion decay, the decay rate of $\left<\theta(\mathbf{r},t)\right>_0$ is just $\Gamma_0$, yielding $\left<\theta(\mathbf{r},t)\right>_0=\theta_0 a^2\delta^{(2)}(\mathbf{r}-\mathbf{r}_0)e^{-\Gamma_0t}$. Therefore, the transition probability reads
\beq
P(T)&=&\int\frac{\mathrm{d}^3\mathbf{k}}{(2\pi)^2}\delta(k_z)\frac{\mathrm{d}^3\mathbf{k}^\prime}{(2\pi)^2}\delta(k^\prime_z)\sum_{\mathrm{polarization}}|G(T;\mathbf{k},\mathbf{k}^\prime)|^2\nn\\
&=&|G_0(T)|^2\frac{\theta_0^2\alpha^2}{4\pi^2}\int\frac{\mathrm{d}^2\mathbf{k}_\parallel}{(2\pi)^2}\frac{\mathrm{d}^2\mathbf{k}_\parallel^\prime}{(2\pi)^2}\sum_{\lambda,\lambda^\prime=1}^2\frac{a^4|f^S_{\lambda\lambda^\prime}(\mathbf{k}_\parallel,\mathbf{k}^\prime_\parallel)|^2}{\Gamma_0^2+(|\mathbf{k}_\parallel|+|\mathbf{k}^\prime_\parallel|)^2},
\eeq
where the integration measure exactly reflects the transverse-ness of the electromagnetic field which constrains $k_z=k^\prime_z=0$ for the emitted photon. The in-plane components of the photon momenta are denoted as $\mathbf{k}_\parallel$ and $\mathbf{k}_\parallel^\prime$. For the generic photon momenta $\mathbf{k}$ and $\mathbf{k}^\prime$ parameterized as
\beq
\mathbf{k}=|\mathbf{k}|(\sin\vartheta\cos\varphi,\sin\vartheta\sin\varphi,\cos\vartheta),~~\mathbf{k}^\prime=|\mathbf{k}^\prime|(\sin\vartheta^\prime\cos\varphi^\prime,\sin\vartheta^\prime\sin\varphi^\prime,\cos\vartheta^\prime),
\eeq
we can adopt the convention of the polarization vectors without losing of generality that
\beq
&&\mathbf{e}_1(\hat{\mathbf{k}})=(\cos\vartheta\cos\varphi,\cos\vartheta\sin\varphi,-\sin\vartheta),~~\mathbf{e}_2(\hat{\mathbf{k}})=(-\sin\varphi,\cos\varphi,0),\nn\\
&&\mathbf{e}_1(\hat{\mathbf{k}}^\prime)=(\cos\vartheta^\prime\cos\varphi^\prime,\cos\vartheta^\prime\sin\varphi^\prime,-\sin\vartheta^\prime),~~\mathbf{e}_2(\hat{\mathbf{k}}^\prime)=(-\sin\varphi^\prime,\cos\varphi^\prime,0).\label{eq:pol}
\eeq
According to this parameterization of the polarization vectors, the in-plane momenta $\mathbf{k}_\parallel$ and $\mathbf{k}_\parallel^\prime$ have $\vartheta=\vartheta^\prime=\pi/2$. Thus, the only non-vanishing form factor is $f^S_{22}(\mathbf{k}_\parallel,\mathbf{k}^\prime_\parallel)=i\sin(\varphi^\prime-\varphi)(|\mathbf{k}^\prime_\parallel|-|\mathbf{k}_\parallel|)/2 $.
Hence, the transition probability is
\beq
P(T)=|G_0(T)|^2\frac{\theta_0^2\alpha^2 a^4}{64\pi^4}\int_0^{\frac{2\pi}{a}}\mathrm{d}|\mathbf{k}_\parallel|\int_0^{\frac{2\pi}{a}}\mathrm{d}|\mathbf{k}^\prime_\parallel|~\frac{|\mathbf{k}_\parallel||\mathbf{k}^\prime_\parallel|(|\mathbf{k}_\parallel|-|\mathbf{k}^\prime_\parallel|)^2}{\Gamma_0^2+(|\mathbf{k}_\parallel|+|\mathbf{k}^\prime_\parallel|)^2}=0.00565|G_0(T)|^2\theta_0^2\alpha^2.
\eeq
Up to the leading order of $\Gamma_0a$, the decay rate and the branching ratio are correspondingly
\beq
\Gamma=\lim_{T\rightarrow+\infty}\frac{P(T)}{T}=0.00565\theta_0^2\alpha^2\Gamma_0,
\eeq
and the braching ratio is $\eta^S=0.00565\theta_0^2\alpha^2\sim 10^{-5}$ for $\theta_0\approx 2\pi$.

The momentum distribution of the emitted photons is determined by
\beq
\int\frac{\mathrm{d}^2\mathbf{k}_\parallel}{(2\pi)^2 }\frac{\mathrm{d}^2\mathbf{k}^\prime_\parallel}{(2\pi)^2}~\tilde{\rho}(\mathbf{k}_\parallel,\mathbf{k}^\prime_\parallel)=\Gamma,
\eeq
yielding
\beq
\tilde{\rho}(\mathbf{k}_\parallel,\mathbf{k}^\prime_\parallel)\propto\sum_{\lambda,\lambda^\prime=1}^2\left|f^S_{\lambda\lambda^\prime}(\mathbf{k}_\parallel,\mathbf{k}^\prime_\parallel)\int \mathrm{d}^2\mathbf{r}\mathrm{d}t~\left<\theta(\mathbf{r},t)\right>_0e^{i(\mathbf{k}_\parallel+\mathbf{k}^\prime_\parallel)\cdot\mathbf{r}-i(|\mathbf{k}_\parallel|+|\mathbf{k}^\prime_\parallel|)t}\right|^2.
\eeq
Denote $\omega=|\mathbf{k}|$ and $\omega^\prime=|\mathbf{k}^\prime|$ and perform the integration over angular variables. For $\left<\theta(\mathbf{r},t)\right>_0=\theta_0a^2\delta^{(2)}(\mathbf{r}-\mathbf{r}_0)$, the energy distribution is
\beq
\rho_0(\omega,\omega^\prime)\propto\frac{(\omega-\omega^{\prime })^2}{(\omega+\omega^\prime)^2+\Gamma_0^2},
\eeq
which does not have a peak. However, in reality $\left<\theta(\mathbf{r},t)\right>_0$ should have a small oscillation, in addition to the exponential decay, due to the energy difference of the initial and the final states. After including this correction, the general form of $\left<\theta(\mathbf{r},t)\right>_0$ reads
\beq
\left<\theta(\mathbf{r},t)\right>_0=\theta_0 a^2\delta^{(2)}(\mathbf{r}-\mathbf{r}_0)e^{-\Gamma_0 t}\left(1+A\cos{\Delta t}\right),\label{eq:aA}
\eeq
where $\Delta$ is the domain wall tension times the axion size reflecting the energy gained from the spin flips, and $A\ll 1$ is the amplitude of the oscillatory part. The energy distribution then reads
\beq
\rho(\omega,\omega^\prime)\propto\rho_{\text{no-peak}}(\omega,\omega^\prime)+\frac{A^2(\omega-\omega^{\prime })^2\left((\omega+\omega^\prime)^2+\Gamma_0^2\right)}{\left((\omega+\omega^\prime)^2+\Gamma_0^2-\Delta^2\right)^2+4\Delta^2\Gamma_0^2},\label{eq:dos}
\eeq
where only the second term contributes to the possible peaks. It can be seen that a sharp peak with peak width $\Gamma_0\ll\Delta$ located at $\omega+\omega^\prime=\Delta$ emerges when $A\neq 0$. In addition, the branching ratio calculated from \Eq{eq:aA} remains the same up to the leading order of $\Delta a$.

\section{Effective electromagnetic action on the surface domain walls}

For a CS-term action on the surface of magnetic topological insulator surfaces $\Sigma$ with the $y$-direction open
\beq
S_\mathrm{CS}=\frac{-i}{4\pi}\int_\Sigma \mathrm{d}^2\mathbf{r}\mathrm{d}t~\epsilon_{\mu\nu\lambda}A_\mu\partial_\nu A_\lambda,
\eeq
its variation under gauge transformation $A_\mu\mapsto A_\mu+\partial_\mu \phi$ is
\beq
S_\mathrm{CS}\mapsto S_\mathrm{CS}-\frac{i}{2\pi}\int \mathrm{d}x\mathrm{d}t~A_x\partial_t\phi-A_0\partial_x\phi=S_\mathrm{CS}-\frac{i}{2\pi}\int \mathrm{d}x\mathrm{d}t~A_x\partial_t\phi.
\eeq
Here in passing to the last equality we adopt the radiation gauge $A_0=0$. Thus, the full theory of the chiral boson field $\phi$ living on the domain wall reads
\beq
S_\mathrm{DW}=\int \mathrm{d}x\mathrm{d}t~\partial_x\phi(\partial_t-v\partial_x)\phi-\frac{i}{2\pi}\int \mathrm{d}x\mathrm{d}t~A_x\partial_t\phi.
\eeq
Upon integrating out the chiral boson field $\phi$ and doing the fourier transformation of the electromagnetic field $A$, the results yields
\beq
S_{\mathrm{DW}}=-\frac{1}{4 \pi} \sum_{\omega, k} A(-\omega,-k) \frac{\omega^2}{k(i \omega-v k)} A(\omega, k),
\eeq
as stated in the main text.

\section{Axion decay from bulk magnon decay}

In the bulk, the axion field is coupled to the electromagnetic field through the $\theta$-term. The total Hamiltonian reads
\beq
&&H=H_0^\mathrm{EM}+H_1[\theta]+H_2^B(t),\nn\\
&&H_2^B(t)=\frac{\alpha}{\pi}\int\mathrm{d}^3\mathbf{r}~\theta(\mathbf{r})\mathbf{E}(\mathbf{r},t)\cdot\mathbf{B}(\mathbf{r},t).\label{eq:Hsf}
\eeq
In parallel to the previous section, the transition amplitude reads
\beq
G(T;\mathbf{k},\mathbf{k}^\prime)=\frac{-i\alpha}{2\pi}G_0(T)f^B(\mathbf{k},\mathbf{k}^\prime)\int \mathrm{d}^3\mathbf{r}\mathrm{d}t~\left<\theta(\mathbf{r},t)\right>_0e^{i(\mathbf{k}+\mathbf{k}^\prime)\cdot\mathbf{r}-i(|\mathbf{k}|+|\mathbf{k}^\prime|)t},
\eeq
where the bulk form factor $f^B(\mathbf{k},\mathbf{k}^\prime)$ (dependence of the polarizations $\lambda$ and $\lambda^\prime$ is implicit for clarity) reads
\beq
f^B(\mathbf{k},\mathbf{k}^\prime)=\frac{1}{2}|\mathbf{k}||\mathbf{k}^\prime|\left(\mathbf{e}^*_\lambda(\hat{\mathbf{k}})\cdot\left(\hat{\mathbf{k}}^\prime\times\mathbf{e}^*_{\lambda^\prime}(\hat{\mathbf{k}}^\prime)\right)+\mathbf{e}^*_{\lambda^\prime}(\hat{\mathbf{k}}^\prime)\cdot\left(\hat{\mathbf{k}}\times\mathbf{e}^*_\lambda(\hat{\mathbf{k}})\right)\right).
\eeq

For the bulk calculation, since the $z$ axis is not equivalent to $x$ and $y$ directions in MnBi$_2$Te$_4$, the dimension of the axion in the $z$ direction should also be in general different to those in the $x$ and $y$ direction. Thus, the axion size should be $a\times a\times w$, where $w\neq a$ in general, and consequently $\left<\theta(\mathbf{r},t)\right>_0$ can be approximated by $\left<\theta(\mathbf{r},t)\right>_0=\theta_0 wa^2\delta^{(3)}(\mathbf{r}-\mathbf{r}_0)$. The transition probability reads
\beq
P(T)&=&\int\frac{\mathrm{d}^3\mathbf{k}}{(2\pi)^3 2|\mathbf{k}|}\frac{\mathrm{d}^3\mathbf{k}^\prime}{(2\pi)^3 2|\mathbf{k}^\prime|}\sum_{\mathrm{polarization}}|G(T;\mathbf{k},\mathbf{k}^\prime)|^2\nn\\
&=&|G_0(T)|^2\frac{\theta_0^2\alpha^2}{4\pi^2}\int\frac{\mathrm{d}^3\mathbf{k}}{(2\pi)^3 2|\mathbf{k}|}\frac{\mathrm{d}^3\mathbf{k}^\prime}{(2\pi)^3 2|\mathbf{k}^\prime|}\sum_{\lambda,\lambda^\prime=1}^2\frac{w^2a^4|f^B_{\lambda\lambda^\prime}(\mathbf{k},\mathbf{k}^\prime)|^2}{\Gamma_0^2+(|\mathbf{k}|+|\mathbf{k}^\prime|)^2}.
\eeq
According to \Eq{eq:pol}, the bulk form factors are parameterized as
\beq
&&f^B_{11}(\mathbf{k},\mathbf{k}^\prime)=-f^B_{22}(\mathbf{k},\mathbf{k}^\prime)=\frac{1}{2}|\mathbf{k}||\mathbf{k}^\prime|(\cos\vartheta-\cos\vartheta^\prime)\sin (\varphi-\varphi^\prime),\nn\\
&&f^B_{12}(\mathbf{k},\mathbf{k}^\prime)=|\mathbf{k}||\mathbf{k}^\prime|\left(\cos\vartheta\cos\vartheta^\prime\cos (\varphi-\varphi^\prime)+\sin\vartheta\sin\vartheta^\prime\right),~~f^B_{21}(\mathbf{k},\mathbf{k}^\prime)=|\mathbf{k}||\mathbf{k}^\prime|\cos (\varphi-\varphi^\prime).
\eeq
We first perform the integration over $\varphi$ and $\varphi^\prime$, as well as the summation over polarizations, which yields
\beq
\int_0^{2\pi}\frac{\mathrm{d}\varphi}{2\pi}\int_0^{2\pi}\frac{\mathrm{d}\varphi^\prime}{2\pi}\sum_{\lambda,\lambda^\prime=1}^2|f^B_{\lambda\lambda^\prime}(\mathbf{k},\mathbf{k}^\prime)|^2=\frac{1}{4}\left(6|\mathbf{k}|^2|\mathbf{k}^\prime|^2-3|\mathbf{k}|^2 k_z^{\prime 2}-2|\mathbf{k}||\mathbf{k}^\prime|k_z k_z^\prime-3|\mathbf{k}^\prime|^2 k_z^2+6k_z^2 k_z^{\prime 2}\right),
\eeq
where $k_z$ and $k^\prime_z$ are the $z$-components of $\mathbf{k}$ and $\mathbf{k^\prime}$, respectively, which have cutoff $2\pi/w$ instead of $2\pi/a$. Upon integrating over $\mathbf{k}$ and $\mathbf{k}^\prime$, we find the bulk branching ratio $\eta^B$ not sensitively dependent on $w/a$, as
\beq
\eta^B=(0.0189\pm 0.0024)\theta_0^2\alpha^2,
\eeq
up to the leading order of $\Gamma_0 w$ and $\Gamma_0 a$. The dependence of $\eta^B/(\theta_0\alpha)^2$ with $\log{w/a}$ is plotted below.

\begin{figure}[htbp]
     \centering
     \includegraphics[width=0.5\linewidth]{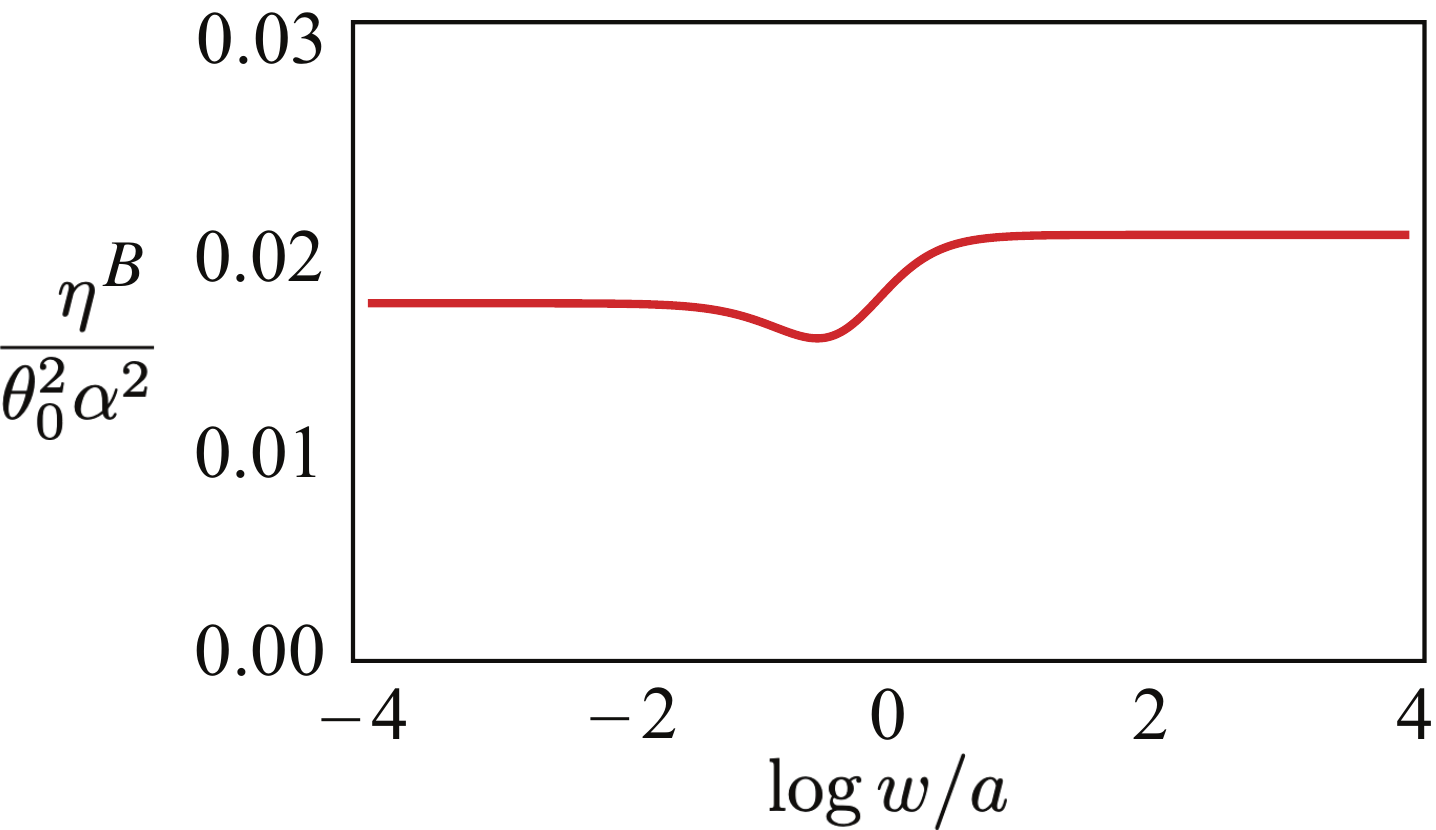}
     \caption{The bulk branching ratio $\eta^B$ as a function of $\log{w/a}$ in the bulk magnon decay setup. It can be concluded that the variation of $w/a$ amounts at most a $15\%$ change in $\eta^B$.}
     \label{fig:branch}    
 \end{figure}

As discussed in the main text, for the bulk magnon $\theta_0$ is at the order of $10^{-2}$. Thus, the branching is $\eta^B\sim 10^{-9}$ which is four orders of magnitude smaller than that in the surface domain wall shrinkage setup.

\section{Field theoretical calculation of bulk axion decay using material details}

The standard field theory describing the coupling between the axion field $\theta(x)$ and the electromagnetic field reads
\beq
\mathcal{L}=m_0^2J\left(\partial_\mu\theta\partial^\mu\theta-m_a^2\theta^2\right)+\frac{\alpha}{\pi}\theta\mathbf{E}\cdot\mathbf{B}\label{eq:L},
\eeq
where the Maxwell term is abbreviated. The period of the axion field $\theta$ is $2\pi$, $m_0$ is the bulk gap of the axion insulator (the time reversal invariant mass), $m_a$ is the axion mass, and $J$ is a dimensionless constant determined from the electronic structure of the axion insulator. By redefining $\theta\mapsto m_0\sqrt{2J}\theta$ and rewriting the electromagnetic field in terms of field strength tensor $F_{\mu\nu}$, the Lagrangian reads
\beq
\mathcal{L}=\frac{1}{2}\left(\partial_\mu\theta\partial^\mu\theta-m_a^2\theta^2\right)+\frac{\alpha}{8\pi m_0\sqrt{2J}}\theta\epsilon^{\mu\nu\rho\lambda}F_{\mu\nu}F_{\rho\lambda},
\eeq
where the periodicity of $\theta$ is now $2\pi m_0\sqrt{2J}$, and $\epsilon^{\mu\nu\rho\lambda}$ is the Levi-Civita symbol. The decay rate of a rest axion into two photons is given by
\beq
\Gamma=\frac{m_a^3c^2}{\pi}\left(\frac{\alpha}{8\pi m_0\sqrt{2J}}\right)^2=\frac{\alpha^2c^2}{128\pi^3}\frac{m_a^3}{Jm_0^2}.
\eeq

The coupling constants $J$ can be computed from the band structure of the axion insulator Mn$_2$Bi$_6$Te$_{11}$, a member of the MnBi$_2$Te$_4$ family. It is described by the Hamiltonian
\beq
&&H(\mathbf{k})=\epsilon(\mathbf{k})+\sum_{i=1}^5R_i(\mathbf{k})\Gamma^i,\nn\\
&&\mathbf{R}(\mathbf{k})=\left(\frac{A_2}{a_2} \sin (k_y a_2),-\frac{A_2}{a_2} \sin (k_x a_2),\frac{A_1}{a_1} \sin (k_z a_1),M(\mathbf{k}),m_5\right)
\eeq
Here 
\beq
&&\epsilon(\mathbf{k})=C+\frac{2D_1}{a_1^2}+\frac{4D_2}{a_2^2}-\frac{2D_1}{a_1^2}\cos (k_za_1)-\frac{2D_2}{a_2^2}\left(\cos (k_xa_2)+
\cos (k_ya_2)\right),\nn\\
&&M(\mathbf{k})=M+\frac{2B_1}{a_1^2}+\frac{4B_2}{a_2^2}-\frac{2B_1}{a_1^2}\cos (k_za_1)-\frac{2B_2}{a_2^2}\left(\cos (k_xa_2)+
\cos (k_ya_2)\right).
\eeq
The five $\Gamma^i$ matrices are mutually anti-commute $\{\Gamma^i,\Gamma^j\}=2\delta^{ij}$. Numerical values of coefficients are
\beq
&&A_1=0.30\text{ eV\AA},~A_2=1.76\text{ eV\AA},~B_1=2.55\text{ eV}\text{\AA}^2,~B_2=14.20\text{ eV}\text{\AA}^2,~M=-0.09\text{ eV},~m_a=1.0\text{ meV},\nn\\
&&a_1=40.91\text{ \AA},~a_2=4.33\text{ \AA},~(\hbar c=1973\text{ eV\AA}).
\eeq
In calculation we use the axion mass measured in experiment $m_a=1.0$ meV instead of the DFT result. Values of $C$ and $D$ are not relevant to us and not listed here. The minus sign of $M$ signals the non-trivial topology of the magnetic insulator Mn$_2$Bi$_6$Te$_{11}$. From this set of coefficients, the bulk gap is found to be $m_0=0.168$ eV. $J$ is given by
\beq
&&J=(\hbar c)^3\int_{\text{BZ}}\frac{\text{d}^3\textbf{k}}{(2\pi)^3}\frac{\sum_{i=1}^4 R_i(\mathbf{k})R_i(\mathbf{k})}{16\left|\mathbf{R}(\mathbf{k})\right|^5}=7.24\times 10^{6}.
\eeq
With numerical values of $m_a$, $J$ and $m_0$, the decay rate yields
\beq
\Gamma=\frac{\alpha^2c^2}{128\pi^3}\frac{m_a^3}{Jm_0^2}=5.27\times 10^{-22}\text{ eV}.
\eeq
For a typical spin relaxation time $\tau_0\approx O(10)$ ms, the branching of the two-photon decay is
\beq
\eta=\frac{\Gamma\tau_0}{\hbar}\approx 10^{-9},
\eeq
which is consistent with the result in the previous section.

\section{Stimulated photon emission}

\subsection{The bulk calculation}

For the stimulated photon emission in the bulk, the transition amplitude is
\beq
G_\lambda(T;\mathbf{k},\mathbf{k}^\prime)=\frac{-i\alpha}{2\pi}G_0(T)f_\lambda(\mathbf{k},\mathbf{k}^\prime)\int \mathrm{d}^3\mathbf{r}\mathrm{d}t~\left<\theta(\mathbf{r},t)\right>_0e^{i(\mathbf{k}-\mathbf{k}^\prime)\cdot\mathbf{r}-i(|\mathbf{k}|-|\mathbf{k}^\prime|)t},
\eeq
where for simplicity we fix the polarization of the stimulating photon to be $\lambda^\prime=2$, and
\beq
f_\lambda(\mathbf{k},\mathbf{k}^\prime)=\frac{1}{2}|\mathbf{k}||\mathbf{k}^\prime|\left(\mathbf{e}^*_\lambda(\hat{\mathbf{k}})\cdot\left(\hat{\mathbf{k}}^\prime\times\mathbf{e}_{2}(\hat{\mathbf{k}}^\prime)\right)+\mathbf{e}^*_2(\hat{\mathbf{k}}^\prime)\cdot\left(\hat{\mathbf{k}}\times\mathbf{e}_{\lambda}(\hat{\mathbf{k}})\right)\right).
\eeq
The transition probability reads
\beq
\tilde{P}(T)=\int\frac{\mathrm{d}^3\mathbf{k}}{(2\pi)^3 2|\mathbf{k}|}\frac{\mathrm{d}^3\mathbf{k}^\prime}{(2\pi)^3 2|\mathbf{k}^\prime|}n(2\pi)^3\delta^{(3)}(\mathbf{k}^\prime-\mathbf{K})\sum_{\lambda}|G_\lambda(T;\mathbf{k},\mathbf{k}^\prime)|^2,
\eeq
where $n$ is the photon density and $\mathbf{K}$ is the momentum of the stimulating photon. By inserting $\left<\theta(\mathbf{r},t)\right>_0$ for bulk ($w=a$ is used for simplicity), the branching of the stimulated emission $\tilde{\eta}^B$ is computed
\beq
\tilde{\eta}^B=0.053\theta_0^2\alpha^2\left( na^3\right)\left(\frac{a}{\Lambda}\right)\approx 10^{-6}\times\frac{Ia^4\theta_0^2}{\hbar c^2},
\eeq
up to the leading order of $\Gamma_0a$ and $a/\Lambda$. Here $\Lambda=2\pi/|\mathbf{K}|$ is the wavelength of the simulating photon, and $I$ is the power density.

\subsection{The surface calculation}

The transition probability reads
\beq
\tilde{P}(T)&=&\int\frac{\mathrm{d}^3\mathbf{k}}{(2\pi)^2}\delta(k_z)\frac{\mathrm{d}^3\mathbf{k}^\prime}{(2\pi)^2}\delta(k_z^\prime)\nu(2\pi)^2\delta^{(2)}(\mathbf{k}^\prime_\parallel-\mathbf{K})|G_{22}(T;\mathbf{k},\mathbf{k}^\prime)|^2\nn\\
&=&|G_0(T)|^2\frac{\nu\theta_0^2\alpha^2 a^4}{16\pi}\int_0^{\frac{2\pi}{a}}\mathrm{d}|\mathbf{k}_\parallel|~\frac{|\mathbf{k}_\parallel|(|\mathbf{k}_\parallel|-|\mathbf{K}|)^2}{\Gamma_0^2+(|\mathbf{k}_\parallel|+|\mathbf{K}|)^2},
\eeq
where in passing to the second line the classical evolution of $\left<\theta(\mathbf{r},t)\right>_0$ and the only non-vanishing form factor $f^S_{22}$ are inserted. $\nu$ is the 2D photon density with unit $[L^{-2}]$. Upon the integration of $\mathbf{k}_\parallel$, the branching is
\beq
\tilde{\eta}^S=0.39\theta_0^2\alpha^2\left(\nu a^2\right)\approx 10^{-5}\times\nu a^2\theta_0^2.
\eeq

